\documentclass{emulateapj}
\usepackage{graphicx,color}
\usepackage{amssymb,amsmath}
\usepackage{appendix}
\usepackage{natbib}
\usepackage{hyperref}
\hypersetup{colorlinks=true,linkcolor=red,urlcolor=cyan,citecolor=blue}
\usepackage[all]{hypcap}

\providecommand{\adsurl}[1]{\href{#1}{ADS}}

\begin{document}
\shorttitle{RELICS: Strong-lensing analysis of two massive clusters}
\shortauthors{Acebron et al.}

\slugcomment{Submitted to the Astrophysical Journal}

\title{RELICS: Strong-lensing analysis of the massive clusters\\ MACS J0308.9+2645 and PLCK G171.9-40.7}
\author{Ana Acebron\altaffilmark{1}*, Nath\'alia Cibirka\altaffilmark{1}, Adi Zitrin\altaffilmark{1}, Dan Coe\altaffilmark{2}, Irene Agulli\altaffilmark{1}, Keren Sharon\altaffilmark{3}, Maru\v{s}a Brada\v{c}\altaffilmark{4}, Brenda Frye\altaffilmark{5}, Rachael C. Livermore\altaffilmark{6}, Guillaume Mahler\altaffilmark{3}, Brett Salmon\altaffilmark{2}, Keiichi Umetsu\altaffilmark{7}, Larry Bradley\altaffilmark{2}, Felipe Andrade-Santos\altaffilmark{8},
Roberto Avila\altaffilmark{2}, Daniela Carrasco\altaffilmark{6}, Catherine Cerny\altaffilmark{3}, Nicole G. Czakon\altaffilmark{7}, William A. Dawson\altaffilmark{9}, Austin T. Hoag\altaffilmark{4}, Kuang-Han Huang\altaffilmark{4}, Traci L. Johnson\altaffilmark{3}, Christine Jones\altaffilmark{8}, Shotaro Kikuchihara\altaffilmark{10}, Daniel Lam\altaffilmark{11}, Lorenzo Lovisari\altaffilmark{8}, Ramesh Mainali\altaffilmark{5}, Pascal A. Oesch\altaffilmark{12}, Sara Ogaz\altaffilmark{2}, Masami Ouchi\altaffilmark{10,13}, Matthew Past\altaffilmark{3}, Rachel Paterno-Mahler\altaffilmark{3}, Avery Peterson\altaffilmark{3}, Russell E. Ryan\altaffilmark{2}, Irene Sendra-Server\altaffilmark{14,15},  Daniel P. Stark\altaffilmark{5}, Victoria Strait\altaffilmark{4}, Sune Toft\altaffilmark{16}, Michele Trenti\altaffilmark{6,17} and Benedetta Vulcani\altaffilmark{6}}
\altaffiltext{1}{Physics Department, Ben-Gurion University of the Negev, P.O. Box 653, Be'er-Sheva 8410501, Israel\\ * anaacebronmunoz@gmail.com}
\altaffiltext{2}{Space Telescope Science Institute, 3700 San Martin Drive, Baltimore, MD 21218, USA}
\altaffiltext{3}{Department of Astronomy, University of Michigan, 1085 South University Ave, Ann Arbor, MI 48109, USA}
\altaffiltext{4}{Department of Physics, University of California, Davis, CA 95616, USA}
\altaffiltext{5}{Department of Astronomy, Steward Observatory, University of Arizona, 933 North Cherry Avenue, Rm N204, Tucson, AZ, 85721, USA}
\altaffiltext{6}{School of Physics, University of Melbourne, VIC 3010, Australia}
\altaffiltext{7}{Institute of Astronomy and Astrophysics, Academia Sinica, PO Box 23-141, Taipei 10617,Taiwan}
\altaffiltext{8}{Harvard-Smithsonian Center for Astrophysics, 60 Garden Street, Cambridge, MA 02138, USA}
\altaffiltext{9}{Lawrence Livermore National Laboratory, P.O. Box 808 L-210, Livermore, CA, 94551, USA}
\altaffiltext{10}{Institute for Cosmic Ray Research, The University of Tokyo,5-1-5 Kashiwanoha, Kashiwa, Chiba 277-8582, Japan}
\altaffiltext{11}{Leiden Observatory, Leiden University, NL-2300 RA Leiden, The Netherlands}
\altaffiltext{12}{Geneva Observatory, University of Geneva, Ch. des Maillettes 51, 1290 Versoix, Switzerland}
\altaffiltext{13}{Kavli Institute for the Physics and Mathematics of the Universe (Kavli IPMU, WPI), The University of Tokyo, Chiba 277-8582, Japan}
\altaffiltext{14}{American School of Warsaw, Warszawska 202, 05-520 Bielawa, Poland}
\altaffiltext{15}{Department of Theoretical Physics, University of Basque
Country UPV/EHU, E-48080 Bilbao, Spain}
\altaffiltext{16}{Cosmic Dawn Center, Niels Bohr Institute, University of Copenhagen, Juliane Maries Vej 30, København, DK-2100, Denmark}
\altaffiltext{17}{Australian Research Council, Centre of Excellence for All Sky Astrophysics in 3 Dimensions (ASTRO 3D)}

\begin{abstract}
Strong gravitational lensing by galaxy clusters has become a powerful tool for probing the high-redshift Universe, magnifying distant and faint background galaxies. Reliable strong lensing (SL) models are crucial for determining the intrinsic properties of distant, magnified sources and for constructing their luminosity function.
We present here the first SL analysis of MACS J0308.9+2645 and PLCK G171.9-40.7, two massive galaxy clusters imaged with the Hubble Space Telescope in the framework of the Reionization Lensing Cluster Survey (RELICS). We use the Light-Traces-Mass modeling technique to uncover sets of multiply imaged galaxies and constrain the mass distribution of the clusters. Our SL analysis reveals that both clusters have particularly large Einstein radii ($\theta_{E}>30\arcsec$ for a source redshift of $z_s=2$), providing fairly large areas with high magnifications, useful for high-redshift galaxy searches ($\sim2$ arcmin$^{2}$ with $\mu>5$ to $\sim1$ arcmin$^{2}$ with $\mu>10$, similar to a typical \textit{Hubble Frontier Fields} cluster). We also find that MACS J0308.9+2645 hosts a promising, apparently bright (J$\sim23.2-24.6$ AB), multiply imaged high-redshift candidate at $z\sim6.4$. These images are amongst the brightest high-redshift candidates found in RELICS. Our mass models, including magnification maps, are made publicly available for the community through the Mikulski Archive for Space Telescopes.  \vspace{0.02cm}
\end{abstract}
\keywords{galaxies: clusters: individual (MACS J0308.9+2645, PLCK G171.9-40.7)--- gravitational lensing: strong}

 \section{Introduction}\label{intro}
 
The study of the early Universe provides essential insight into the mechanisms of galaxy formation and evolution over cosmic time, as well as the role played by primeval galaxies in the reionization of the Universe \citep{Loeb2013}. Measuring the number counts of high-redshift galaxies allows us to derive their luminosity function (LF) and its evolution with redshift \citep[e.g.,][]{Trenti2010, Oesch2014, McLeod2016, Livermore2017, Bouwens2017} which are extensively used to describe the properties of these sources in a statistical way. Particularly, by building up the LF in the rest-frame ultra-violet (UV), it is possible to infer the cosmic star formation rate density at early times and assess the role of early galaxies during reionization \citep{Bunker2004, Bouwens2010, Ellis2013, Finkelstein2015}. 

This field faces, however, important observational challenges associated with the reliable detection of distant, low-luminosity galaxies. Massive strong gravitational lenses -- such as galaxy clusters -- act as natural telescopes, allowing us to peer deeper into the Universe and infer the properties of background sources that are below the resolution or sensitivity limit of blank fields at the same depth. While blank field surveys can provide robust constraints on the bright-end of the LF \citep{McLure2013, Bouwens2014}, strong lensing (SL) clusters can, in a complementary way, constrain the fainter end by magnifying background sources that would otherwise be too faint to be detected \citep{richard2008, Zheng2012, Coe2013,  Zitrin2014, Atek2015, Livermore2017,Bouwens2017b}.

SL by galaxy clusters therefore, thanks to their magnification power, allows us to characterize background sources in unprecedented detail, but also entails a reduced source-plane area, so that strong constraints on the faint-end of the high-redshift UV LF are difficult to obtain \citep{Atek2015,Bouwens2017,Atek2018}. The net effect of this \textit{magnification bias}, i.e., the trade-off between the two competing effects of seeing intrinsically fainter objects but in a smaller probed area (i.e., the area benefiting from high-magnifications is correspondingly much smaller), depends on the shape of the high-redshift LF, but generally (and assuming typical LFs), we expect to observe more galaxies thanks to lensing, especially at the apparently brighter end \citep{Turner1984,Broadhurst1995MB,Wyithe2011}.

The recent \textit{Hubble Space Telescope} (HST) Treasury program RELICS\footnote{\url{https://relics.stsci.edu/}}\citep[PI: D. Coe; Coe et al., in preparation;][]{Cerny2017, Salmon2017} aims to uncover a substantial sample of high-redshift galaxies, by combining HST observations with the magnification power from a large sample of massive galaxy clusters, thus minimizing uncertainties due to cosmic variance.
In order to reliably determine the intrinsic properties of background sources, construct their LF, and probe their role in reionization, SL models -- especially the derived magnification maps -- of these foreground lenses, are then crucial.

\begin{table*}
	\caption{Properties of the RELICS clusters analyzed in this work}            
	\label{table:1}      
	\centering  
        {\renewcommand{\arraystretch}{1.1}
	\begin{tabular}{c c c c c c c}        
		\hline\hline                 
		Cluster & R.A. & Dec& Redshift & scale & Planck SZ mass\tablenotemark{a} & Planck SZ rank\tablenotemark{b} \\  
		&[J2000]&[J2000]&& [kpc/"]& [$10^{14}M_\odot$]& \\ 
		\hline  \vspace{0.2cm}        
		MACSJ0308.9+2645 & 03:08:59 & +26:45:30 &0.356& 4.994 & 10.76& 12\\
		PLCK G171.9-40.7 & 03:12:57&+8:22:19 &0.270& 4.135&10.71 &16\\ 
		\hline\hline                                
	\end{tabular}}
         \tablecomments{}
      \tablenotetext{1}{The mass estimate corresponds to M500 from \citet{Planck2015}.}
      \tablenotetext{2}{The SZ mass ranking in the Planck cluster catalog PSZ2.}
\end{table*}

Assuming a mean background density of sources, larger strong lenses, i.e., those with a larger critical area or Einstein radii, should comprise on average more lensed sources and are thus considered particularly useful lenses. In fact, since these massive clusters sit at the high end of the cosmic mass function, they are also useful for constraining cluster physics \citep{Rasia2013,Ettori2015}, structure evolution, and cosmological models \citep{Blanchard1998,Planck2014}. However, given the shape of the mass function, more massive lenses are scarcer, with only about a dozen clusters known with effective Einstein radii above 30\arcsec \citep{Broadhurst2005b,Richard2010,Zitrin2017}.\\
In this work we present the first SL analysis of two massive galaxy clusters from the RELICS sample, MACS J0308.9+2645 and PLCK G171.9-40.7, performed with our Light-Traces-Mass method \citep[LTM; e.g., ][]{zitrin2015}. \\ 
This paper is organized as follows. Section \ref{sec:data} presents the data and observations, Section \ref{sec:lens_model} describes the adopted SL modeling technique as well as our best-fit models. Our findings are presented and discussed in Section \ref{sec:results}. Finally, our conclusions are summarized in Section \ref{sec:summary}. \\
Throughout this work, we adopt the standard ${\Lambda\mathrm{CDM}}$ flat cosmological model with the Hubble constant $H_0 = 70$ $   \mathrm{km~s^{-1}~Mpc^{-1}}$ and $\mathrm{\Omega_{M}}=0.3$. Magnitudes are quoted in the AB system. Errors are typically 1$\sigma$ unless otherwise noted.

\section{Target, data and observations} \label{sec:data}
The clusters analyzed in the present work are part of the RELICS cluster sample. The RELICS project aims at analyzing 41 massive clusters in order to efficiently search for and study magnified high-redshift galaxies. RELICS targeted 21 among the 34 most massive clusters according to their Sunyaev Zel'dovich  \citep[SZ,][]{Sunyaev1972} mass estimates in the Planck catalogue \citep{Planck2016}, that lacked HST/IR imaging. In order to maximize exceptional lenses in the RELICS sample, the other half of the sample was selected based on several criteria, such as mass estimations from X-ray \citep[MCXC,][]{piffaretti2011,mantz2010}, weak lensing \citep{sereno2014,applegate2014,vonderlinden2014,umetsu2014,hoekstra2015} and SZ -- especially from South Pole Telescope \citep[SPT,][]{Bleem2014} and Atacama Cosmology Telescope \citep[ACT,][]{Hasselfield2013} data -- as well as lensing-strength predictions for some Sloan Digital Sky Survey clusters \citep[SDSS,][]{wong2013,wen2012}. Each cluster in the sample was then observed for a total of 3 orbits with the \textit{Advanced Camera Survey} (ACS- F435W, F606W, F814W) and 2 orbits with  the \textit{Wide Field Camera 3} (WFC3/IR- F105W, F125W, F140W, F160W) except for cases where HST/ACS archival data were already available.

As detailed in  \citet{Cerny2017}, the RELICS team has delivered reduced HST images and photometric source catalogues for the clusters. The photometry is measured with isophotal apertures by SExtractor \citep{Bertin1996} in dual-image mode, based on the final drizzled 0.06" images, and Bayesian photometric redshifts (hereafter $\mathrm{z_{phot}}$) are then derived using the \textit{Bayesian Photometric Redshift} program \citep[BPZ,][]{Benitez2000, Benitez2004, Coe2006} from seven HST band imaging-data (from RELICS observations and HST archival data). These data products are available for the community through the Mikulski Archive for Space Telescopes (MAST)\footnote{\url{https://archive.stsci.edu/prepds/relics/}\label{mast}}.

In this paper we perform a SL analysis of MACSJ0308.9+2645 (MACS0308 hereafter) and PLCKESZ G171.94-40.65 (PLCK G171.9 hereafter). Some relevant details for the clusters are presented in \autoref{table:1}.

\begin{figure*}
	\centering
	\includegraphics[width=0.76\linewidth]{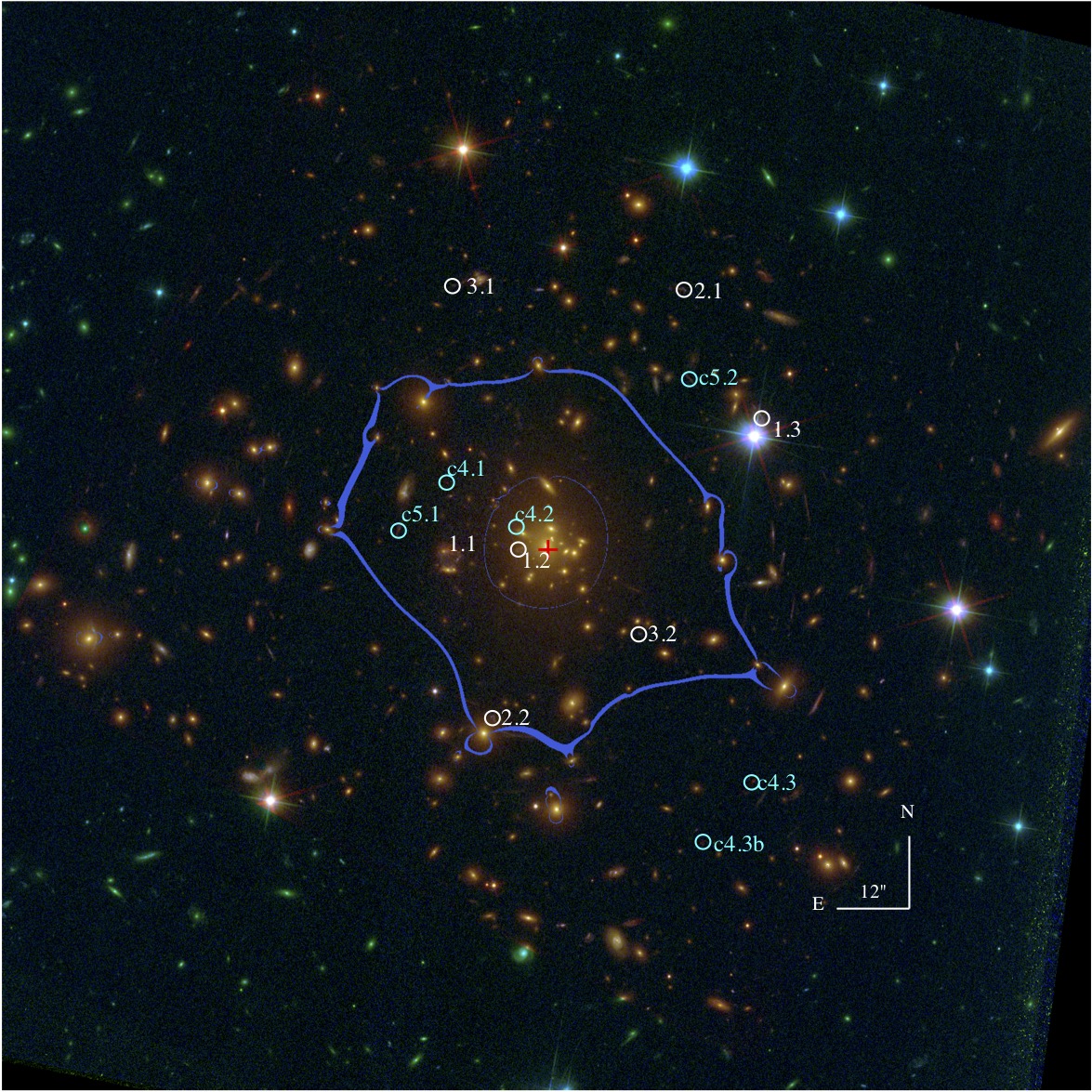} 
    \caption{Color-composite image of MACS0308. Image was created using the HST/ACS passbands F435W (blue), a combination of F606W+F814W (green), and a combination of the HST/WFC3IR passbands F105W+F125W+F140W+F160W (red). The critical curves from our best-fit model are displayed in violet for a source at $z\sim 1.15$. The BCG position is indicated with a red cross. Multiple images used as constraints are labelled according to \autoref{table:0308}. Less secure multiple-image candidates are indicated in cyan, and are not used as constraints in the modeling.}
	\label{macs0308cc}
\end{figure*}

MACSJ0308 ($z=0.35$), displayed in \autoref{macs0308cc}, is part of the X-ray luminous \textit{Massive  Cluster  Survey}  (MACS) cluster sample \citep{Ebeling2001}. MACS0308 had also been previously observed with the ACS for 0.5 orbit in the F606W band, and 0.5 orbit in F814W (HST GO programs 12166, 12884, PI: Ebeling). While the morphology from the X-ray emission map suggests a relaxed and regular cluster, its temperature distribution reveals the presence of two cool cores, possibly associated with an ongoing merger of two substructures \citep{parekh2017}. A diffuse radio emission was detected as well between these substructures, extending towards the N-S direction.

\begin{table*}
	\caption{Multiple images and candidates for MACS0308.}       
	\label{table:0308}      
	\centering                          
	\begin{tabular}{c c c c c c c}        
		\hline\hline                 
		Arc ID & R.A. & Dec& $\mathrm{z_{phot}}$ [$\mathrm{z_{min}}$-$\mathrm{z_{max}}$]\tablenotemark{a} & $\mathrm{z_{model}}$[$95\%$ C.I.]\tablenotemark{b} & Comments & individual RMS (")\tablenotemark{c}\\  
		&[J2000]&[J2000]&&  & \\
		\hline          
		1.1   & 03:08:57.171 & +26:45:37.22  & 1.15 [1.12-1.22] &1.15\tablenotemark{d} &  & 0.9\\  
		1.2   & 03:08:56.261 & +26:45:37.22 & - &"&in BCG's light  & 0.8\\  
		1.3   & 03:08:53.345 & +26:45:57.32  & -&"& behind a star & 0.5\\  
		\hline
		2.1   & 03:08:54.266 & +26:46:20.15 &1.41 [1.17-2.24]  &1.97 [1.80-2.09]& & 1.1\\  
		2.2   & 03:08:56.611 & +26:45:09.16 & 1.93 [1.83-2.10]&"& & 0.4\\  
		\hline
		3.1   & 03:08:57.107 & +26:46:20.97  & --&1.73 [1.43-1.80]& & 0.8\\  
		3.2   & 03:08:54.810 & +26:45:23.39  &1.12 [1.09-2.41] &" & & 0.9\\  
        \hline
		c4.1   & 03:08:57.179 & +26:45:48.34  & 6.42 [6.06-6.96]&$\sim6.4$& not used as constraint& -\\  
		c4.2   & 03:08:56.347 & +26:45:41.59  &- & "& " $\&$ in BCG's light& -\\ 
        c4.3   & 03:08:53.407 & +26:44:58.93  &6.27 [6.14-6.37] & "& "& -\\ 
       c4.3b   & 03:08:54.027 & +26:44:48.71  &2.02 [1.67-2.33] & "& "& -\\ 
         \hline
		c5.1   & 03:08:57.786 & +26:45:40.55  & 1.12 [1.07-1.37]&$\sim 1.24$& not used as constraint& -\\  
		c5.2   & 03:08:54.205 & +26:46:05.54  &1.22 [0.98-1.71] &" & "& -\\ 
		\hline\hline 
	\end{tabular}
    \tablecomments{}
  \tablenotetext{1}{Photometric redshift with upper and lower limits, based on the BPZ estimates from RELICS catalogue with the $95\%$ confidence range. - indicates an image for which its $\mathrm{z_{phot}}$ could not be measured due to light contamination or poor signal-to-ratio.}
  \tablenotetext{2}{Redshift prediction based on our best-fit model.}
 \tablenotetext{3}{RMS between the observed and model-predicted multiple images from our best-fit model.}
 \tablenotetext{4}{fixed redshift.}
\end{table*}

\begin{figure*}
	\centering
	\includegraphics[width=0.76\linewidth]{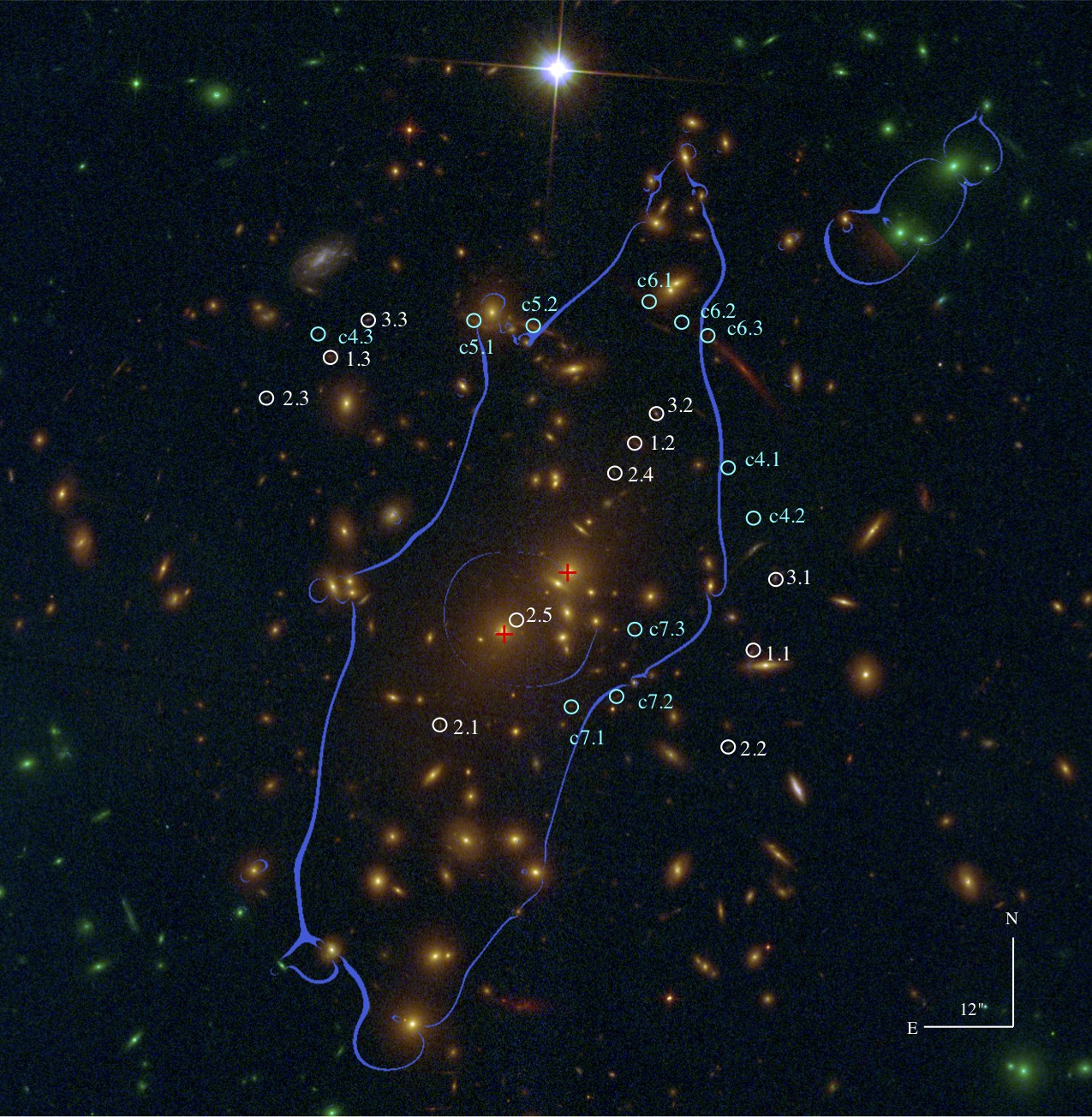} 
	\caption{Color-composite image of PLCK G171.9. Image was created using the HST/ACS passbands F435W (blue), a combination of F606W+F814W (green), and a combination of the HST/WFC3IR passbands F105W+F125W+F140W+F160W (red). Critical curves from our best-fit model are displayed in violet for a source at $z\sim 2.0$. The two BCG positions are indicated as red crosses. Multiple images used as constraints are labelled according to \autoref{table:171}. Less secure multiple-image candidates are indicated in cyan, and are not used as constraints in the SL modeling.}
	\label{plck171cc}
\end{figure*} 

\begin{table*}
	\caption{Multiple images and candidates for PLCK G171.9 }            
	\label{table:171}      
	\centering                          
	\begin{tabular}{c c c c c c c}        
		\hline\hline                 
		Arc ID & R.A. & Dec& $\mathrm{z_{phot}}$ [$\mathrm{z_{min}}$-$\mathrm{z_{max}}$]\tablenotemark{a} & $\mathrm{z_{model}}$[$95\%$ C.I.]\tablenotemark{b} & Comments & individual RMS (")\tablenotemark{c}\\  
		&[J2000]&[J2000]&&  & \\
		\hline          
		1.1   & 03:12:55.230&+08:22:08.10&  2.00 [1.90-2.08]& 2.00\tablenotemark{d}& & 1.7\\  
		1.2   & 03:12:56.281&+08:22:36.36 &1.98 [1.87-2.10] & "& & 1.9\\  
		1.3   & 03:12:59.036&+08:22:47.85 & 1.41 [1.34-2.18]& "&&0.4\\  
		\hline
		2.1   & 03:12:58.043&+08:21:58.16& 0.08 [0.07-0.90]& 2.10 [2.05-2.58]&& 3.0\\  
		2.2   & 03:12:55.447&+08:21:55.38 & 3.34 [0.30-3.50]& "& &2.8\\  
		2.3   & 03:12:59.630&+08:22:42.42 & 3.40 [0.35-3.51]& "&& 4.1\\ 
		2.4   & 03:12:56.467&+08:22:32.21& 0.3 [0.10-3.79] & "& &0.6\\  
		2.5   & 03:12:57.387&+08:22:12.06& - & "& in BCG's light&1.2\\ 
		\hline
		3.1   & 03:12:55.015&+08:22:18.30& 1.99 [1.89-2.16] & 2.38 [2.24-2.72]&& 0.7\\  
		3.2   & 03:12:56.099&+08:22:40.68& 2.39 [1.43-2.46] & "& &0.9\\  
		3.3   & 03:12:58.699&+08:22:53.04& 1.41 [1.40-2.18]& "& &1.0\\  
		\hline
		c4.1   & 03:12:55.503 & +08:22:34.21 & 2.86 [2.15-3.24]&$\sim3.8$ & not used as constraint&-\\   
		c4.2   & 03:12:55.171 & +08:22:26.60 & -&- & ", Sys. can also be high-z&-\\ 
        c4.3   &  03:12:59.089& +08:22:50.69 & -&- & ", (see text)&-\\ 
        \hline
		c5.1   &  03:12:57.714& +08:22:52.60 & -&$\sim0.86$& not used as constraint&-\\   
		c5.2   &  03:12:57.187& +08:22:51.62 & 0.84 [0.62-0.90]&- & "&-\\ 
       \hline
		c6.1   & 03:12:55.727 & +08:22:51.88 & -&$\sim0.97$ & not used as constraint&-\\   
		c6.2   & 03:12:55.970 & +08:22:53.43 & -&- & "&-\\ 
        c6.3   & 03:12:56.254 & +08:22:55.95 & -&- & "&-\\ 
        \hline
       c7.1  & 3:12:56.871 & +8:22:01.05 & 1.22 [1.15-2.36]& $\sim1.2$  & not used as constraint&-\\ 
       c7.2  & 3:12:56.436 & +8:22:02.30 & 2.34 [2.22-2.63]&- & "&-\\
     c7.3  & 3:12:56.290 & +8:22:11.44 &  0.72 [0.11-0.71]&- & "&-\\
	\hline\hline                                
	\end{tabular}
     \tablecomments{}
      \tablenotetext{1}{Photometric redshift with upper and lower limits, based on the BPZ estimates from RELICS catalogue with the $95\%$ confidence range. - indicates an image for which its $\mathrm{z_{phot}}$ could not be measured due to light contamination or poor signal-to-ratio.}
  \tablenotetext{2}{Redshift prediction based on our best-fit model.}
 \tablenotetext{3}{RMS between the observed and model-predicted multiple images from our best-fit model.}
 \tablenotetext{4}{fixed redshift.}
\end{table*}

The other cluster analyzed in this work is PLCK G171.9 ($z=0.27$), presented in \autoref{plck171cc}. PLCK G171.9 was discovered by Planck observations through the SZ effect with a SNR $\sim 10.7$ and confirmed by \textit{XMM-Newton} X-ray follow-up which showed that the galactic overdensity was coincident with the X-ray luminosity peak, with the BCG's position only slightly offset from the X-ray peak (\citealt{Planck2011, Planck2012}). \citet{Giacintucci2013} reported the discovery of a radio halo, often associated with cluster mergers. Indeed, their X-ray analysis of the cluster revealed an asymmetric surface brightness and temperature distribution, which may indicate a disturbed Inter Cluster Medium (ICM) consistent with a recent cluster merger scenario along the NW-SE axis.

\section{Lens modeling} \label{sec:lens_model}
We perform the SL analysis using the LTM method by \citet[][see also \citealt{broadhurst2005,zitrin2015}]{zitrin2009}. We give here a brief overview of the pipeline. The LTM method is based on the assumption that the general dark matter (DM) distribution in the cluster is traced by the distribution of cluster galaxies, where the mass of each cluster galaxy is generally proportional to its luminosity. 

The starting point of the modeling is therefore the identification of cluster members, following the red-sequence method \citep{Gladders2000}. The magnitudes measured from the F606W and F814W filters are used to draw a color-magnitude diagram, choosing galaxies down to 23 AB within $\pm0.3$ mag of this sequence, a value often used in the literature \citep[e.g.,][]{DeLucia2008}. Stars are excluded from our selection primarily by applying a cut-off value for the \textit{stellarity} index of $<0.95$, and only considering objects with magnitudes fainter than 17 AB. We also rely in part on the help of a size-magnitude relation, plotting the FWHM versus the F814W magnitude, to help identify stars which may have been selected.
A subsequent visual inspection is performed, discarding further interloping galaxies or artifacts, and including high-probability cluster galaxies that were missed in the initial selection.
A symmetric power-law surface mass-density distribution, scaling linearly in amplitude with luminosity, is then assigned to each galaxy defined as:
\begin{equation}
\label{eqdis}
\Sigma(r)=Kr^{-q}\mathrm{,}
\end{equation}
where r is the galaxy's radius. All galaxies have the same power-law exponent, $q$, which is a free parameter of the model, and $K$ embeds the linear scaling with the galaxy's measured flux. By integrating \autoref{eqdis}, the enclosed mass for each galaxy is then given by:
\begin{equation}
M(<\theta) = \frac{2 \pi K}{2-q}(D_L \theta)^{2-q},
\end{equation}
where $\theta$ is the angular position and $\mathrm{D_L}$ is the angular diameter distance to the lens. The sum of all galaxy mass distributions then defines the contribution of the galaxy component of the model. \\
The deflection angle due to each galaxy is then given by: 
\begin{equation}
\alpha_{gal}(\theta)=\dfrac{4GM(<\theta)}{c^2\theta}\dfrac{D_{LS}}{D_S D_L}\mathrm{,}
\end{equation}
where the galaxy's enclosed mass $M(<\theta) \propto \theta^{2-q}$ is proportional to its luminosity. $\mathrm{D_S}$ and $\mathrm{D_{LS}}$ are the angular diameter distances to the source, and between the lens and the source, respectively. The galaxy's deflection angle can then be rewritten as: 
\begin{equation}
\alpha_{gal}(\theta)=K_qF\theta^{1-q}\mathrm{,}
\end{equation}
where $K_{q}$ is proportional to the lensing distance ratio $\mathrm{D_{LS}}/\mathrm{D_{S}}$ and represents the overall normalization. 

Since we expect the DM distribution to be smoother, the co-added galaxy distribution is then smoothed with a 2D Gaussian whose width, $S$, is the second free parameter of the model. From this smooth mass-density map the deflection field for the DM component of the model, $\vec{\alpha}_{DM}(\vec{\theta})$, is calculated. The third free parameter is then the scaling of this smooth DM component relative to the total galaxy component, which we denote as $K_{gal}$. The fourth free parameter is the overall normalization, $K_{q}$.
Since galaxies and the underlying dark matter distribution are not expected to trace each other rigorously, a two-parameter external shear (introducing a large-scale ellipticity to the magnification map) is also added to allow further flexibility. The external shear is parametrized by its amplitude $\gamma$ and its position angle $\phi$. The total deflection field $\vec{\alpha}_T(\vec{\theta})$, is obtained by adding the contribution of the different components considered in the model:
\begin{equation}
\vec{\alpha}_T(\vec{\theta})=K_{gal}\vec{\alpha}_{gal}(\vec{\theta})+ (1-K_{gal})\vec{\alpha}_{DM}(\vec{\theta})+\vec{\alpha}_{ex}(\vec{\theta})\mathrm{,}
\end{equation}
so that the galaxy and the DM components are scaled by the factor $K_{gal}$ and $(1-K_{gal})$, respectively. 

In order to further improve the fit, the relative
weight of the brightest
cluster galaxies (BCGs) can be freely optimized in the
minimization procedure, i.e., they can be allowed to deviate from the M/L scaling relation assigned to all other cluster members \citep[see also][]{VonderLinden2007,Kormendy2013,Newman2013}. In addition, a core and ellipticity can also be introduced for the BCGs, whose parameters -- core radius, ellipticity and position angle, respectively -- can also be freely optimized in the minimization procedure, adding more degrees of freedom to the model.

The position and source redshift (where available) of multiple image families are used as constraints for the SL modeling. The goodness of fit is assessed using a $\chi^2$ criterion quantifying the reproduction of multiple-image positions in the image plane (we assume a positional uncertainty of $0.5"$ for the multiple images), written as: 
	\begin{equation}
	\centering
	\chi^2 = \sum\limits_{i=1}^{n} \dfrac{( x_i^{pred} - x_i^{obs})^2 + ( y_i^{pred} - y_i^{obs})^2}{\sigma_{i}^2} \mathrm{,}
	\end{equation} 
with $\mathrm{x_i^{obs}}$, $\mathrm{y_i^{obs}}$ and $\mathrm{x_i^{pred}}$, $\mathrm{y_i^{pred}}$ being the observed and model-predicted positions of the multiple images, respectively, and $\sigma_{i}$ the corresponding positional uncertainty.\\
The optimization of each model is carried out with several
thousand Monte Carlo Markov Chain (MCMC) steps. The goodness-of-fit of a model can also be assessed with the root-mean-square (RMS) between the observed and model-predicted positions of the multiple images in the image plane, which is computed as follows: 
\begin{equation}
RMS = \sqrt[]{\dfrac{1}{N_{img}} \sum_{i=1}^{n} \left(( x_i^{pred} - x_i^{obs})^2 + ( y_i^{pred} - y_i^{obs})^2 \right)} \mathrm{,}
\end{equation}
with $\mathrm{N_{img}}$ being the total number of images.\\
Being strongly coupled to the light distribution, so that a minimum of free parameters are needed to generate a mass model while having sufficient flexibility,  LTM is a powerful method to both identify new multiple images, and constrain the cluster mass distribution \citep[e.g.,][see also \citealt{broadhurst2005}]{zitrin2015,Zitrin2017}. Our SL models typically include about 10-15 total free parameters when accounting, in addition, for freely optimized BCGs and source redshifts where needed.


The LTM method allows to iteratively predict the appearance and orientation of multiple images. Using an initial model constructed by adopting typical parameter values for example, a typical $q=1.3$ for the power law, or a $\sim15$\% galaxies-to-DM relative weight, we iteratively delens-relens (i.e., send to the source plane and back to the image plane by using the lens equation), multiple image candidates, in particular distinct-looking or blue arclets, and look in the data for potential, similar-looking counter images where the initial model predicts them. Multiple-image families are eventually identified -- guided by the model's prediction -- by their colors, morphology and symmetry. This process is repeated iteratively as our models are refined using the identified systems. For the modeling we only use as constraints the position of multiply imaged systems that we consider secure, but also present here other possible multiple-image candidates. While somewhat subjective, by secure systems we refer to those whose agreement with the model prediction, internal details, similar colors, and symmetry, leave essentially no doubt these are images of the same source.

All multiple images presented here are, to our knowledge, the first published for both clusters, and we found no record of spectroscopic redshifts available for any of the systems considered here. Therefore, for each cluster, we fix the redshift of the system with the more reliable photometric estimation to its mean redshift estimate by BPZ, and leave the redshift of other systems as free parameters to be optimized in the minimization procedure (allowing the corresponding $\mathrm{D_{LS}}/\mathrm{D_S}$ ratio for each system to vary by $\pm0.2$). The implications of such assumptions regarding the accuracy of our SL models \citep[see also][]{Cerny2017} are further discussed in Section \ref{sec:results}.

\subsection{MACS J0308.9+2645}
In MACS0308 we identified three multiple-image systems, displayed in \autoref{macs0308cc} and listed in \autoref{table:0308}. The first system refers to a spiral galaxy with three multiple images. The third image of this system, 1.3, lies partially behind a star where the LTM predicts it, and can be more easily identified in the WFC3/IR images.
The second system, comprising two multiple images, has a particular shape that allows for a reliable identification (see \autoref{stamps0308}). The third system is a faint blue galaxy with two multiple images, labelled 3.1 and 3.2. Our model predicts a third, fainter counter-image near the center of the cluster but, likely due to its faintness and contamination by the BCG's light, we do not detect it. The reproduction of multiple images identified in this cluster, by our best-fit model, is shown in \autoref{stamps171}.

Our methodology allows us to predict other sets of multiple images, that we consider less secure and thus do not use them as constraints for our modeling. These candidate systems are designated as such in \autoref{table:0308} and \autoref{macs0308cc}. Candidate system c4 is a drop-out system with a $z_{phot}\sim6.4$, and geometrically predicted to be at high redshift by our model in agreement with its photometric redshift. This system is amongst the brightest high-z candidates found in RELICS, and is further discussed in Section \ref{sec:results}. The candidate multiple images of system c5 present similar colors implying they may be related, and our SL best-fit model predicts a $\mathrm{z_{model}}\sim1.24$ for this scenario.
In addition to these candidate systems, we note that there are many other elongated, lensed arclets seen in \autoref{macs0308cc}, some of which may be in principle multiply imaged, although a lack of internal details and distinct appearance challenge their identification as such. Future dedicated efforts including spectroscopic redshift measurements can help find additional systems in this rich cluster field.

We model MACS0308 leaving as a free parameter the weight of the central BCG. Given its apparent round shape, we assign no ellipticity to the BCG, but allow for a core with a radius that can reach values of up to $120$ kpc. The main source redshift to which we scale our model is set to the photometric redshift of system 1 (with a reliable estimate from BPZ). The redshifts for systems 2 and 3 are left as free parameters of the model and optimized by the SL MCMC modeling pipeline. 
The critical curves (for a source at $z_s = 1.15$) for our final best-fit model, which has an image reproduction $RMS = 0.8\arcsec$, are shown in \autoref{macs0308cc}.

\subsection{PLCK G171.9-40.7}
The model for PLCK G171.9 is based on the identification of three multiple-image systems, shown in \autoref{plck171cc}. Their properties are detailed in \autoref{table:171}. The first and third systems comprise three images each. The second source is multiply imaged into five images. Our model does not predict any further counter-images for these systems. \\
We find that in our modeling framework, the fit for PLCK G171.9 is improved by allowing the weights of the two central BCGs to vary during the optimization. The ellipticity of the first BCG is left as a free parameter and can vary up to $0.4$, whereas the second BCG is also assigned with an ellipticity, set to the value measured for it by SExtractor. The core radii of the two BCGs are also left free and can reach values up to $100$ kpc. The main source redshift we used for this cluster was set to the photometric redshift of system 1, while the redshift for the other two systems are again left free to be optimized in the minimization.
The final best model has an image reproduction $RMS=2.0\arcsec$ and its resulting critical curves (for a source at $z_s = 2.0$) are shown in \autoref{plck171cc}. The reproduction of multiple images is seen in \autoref{stamps171}. We also identify other sets of candidate multiple images (considered less secured), listed in \autoref{table:171} and displayed in \autoref{plck171cc}. 
Two images of System c4 form a thin and faint red arc (as seen in the composite ACS/WFC3IR image) straddling the critical curves (images c4.1 and c4.2). Our best-fit SL model predicts a third counter image at the other side of the cluster, c4.3, and implies a redshift of $\mathrm{z_{model}\sim 3.8}$ for this system (however, given the lensing distances involved, it is only possible to determine geometrically from lensing that the redshift is larger than $\gtrsim3.5$; in fact, we note that some of the models we probed prefer a higher-redshift solution of $z\sim6-7$). Given its faintness, a robust $\mathrm{z_{phot}}$ estimation for this system is challenging.

The images that we denote c5.1 and c5.2 are similar looking and appear to be lensed, lying next to a few cluster galaxies. However, it is currently unclear whether the two images constitute counter images of the same source galaxy (with a $\mathrm{z_{model}}\sim1.5$), in which case a third counter image is expected on the other side of the cluster and which we do not clearly identify; one main image and a combination of partial counter images locally lensed by the adjacent cluster galaxies ($\mathrm{z_{model}}\sim0.86$); or images of two different (but possibly related) galaxies at a somewhat lower redshift. Further detailed examination will be needed to determine the underlying scenario, which we leave for future work.\\ 
Candidate system c6 appears as a faint red arc, lying just below a cluster galaxy which evidently contributes to its lensing. The c6 arc might be related to the brighter lensed red arc next to it at R.A.=03:12:55.30, Dec=+08:22:46.31 (a scenario for which we obtain $z_{model}\sim0.97$). However, also here the picture is still unclear, and the faint arc, labelled as c6.1-c6.2, might only be locally lensed and not at all related to c6.3 (next to the bright arc), a scenario for which our SL model predicts a lower redshift of $z_{model}\sim0.7$.
Finally, we report candidate system c7, composed of two multiple images whose shape and colors are similar, and their parity agrees well with the expected symmetry. Our SL model predicts additional counter-images for this candidate system, one of them we tentatively identify in the data, presented in \autoref{table:171}, although another close-by image (located at R.A.=03:12:56.273, Dec=+08:22:14.29) could also be the actual counterpart. Additional images of this system are predicted in the cluster's center and in the north-east region, which we do not detect, possibly due to light contamination and expected faintness.

\begin{figure*}
	\centering
	\includegraphics[width=0.53\linewidth]{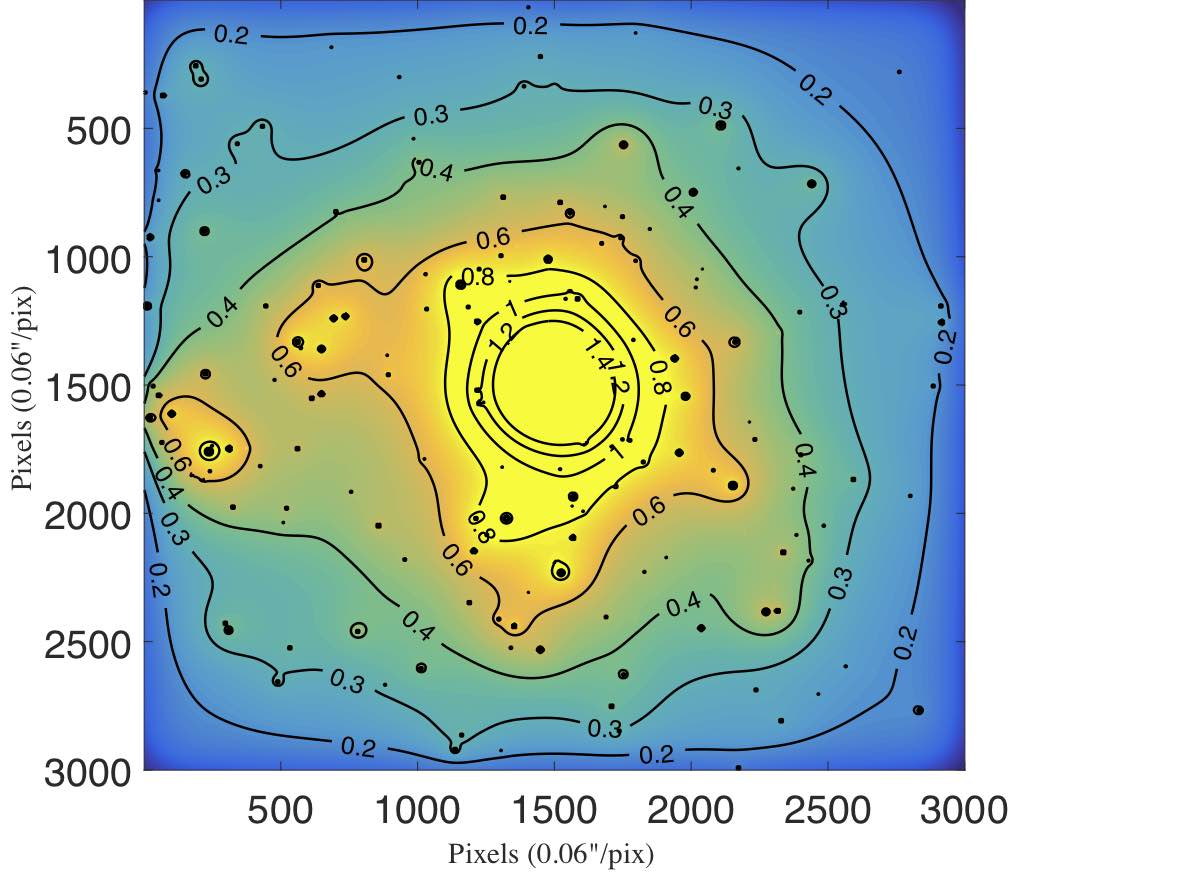} \hfill
 	\includegraphics[width=0.4475\linewidth]{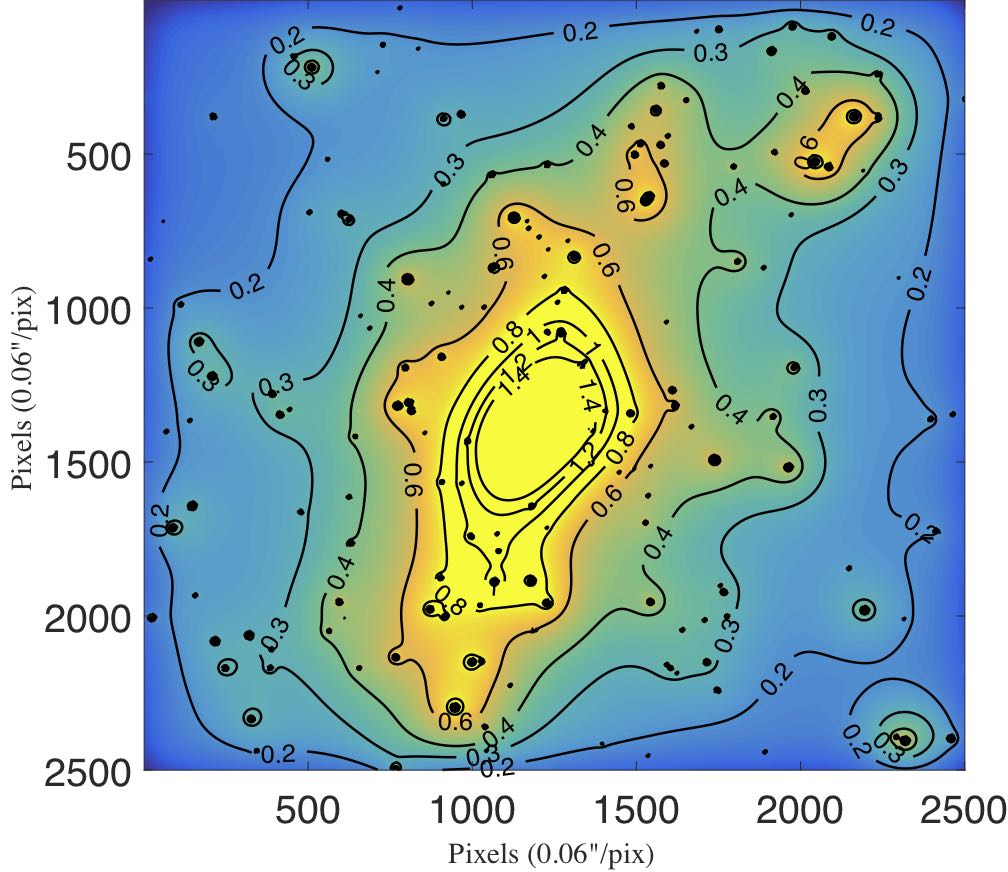} \\

\includegraphics[width=0.47\linewidth]{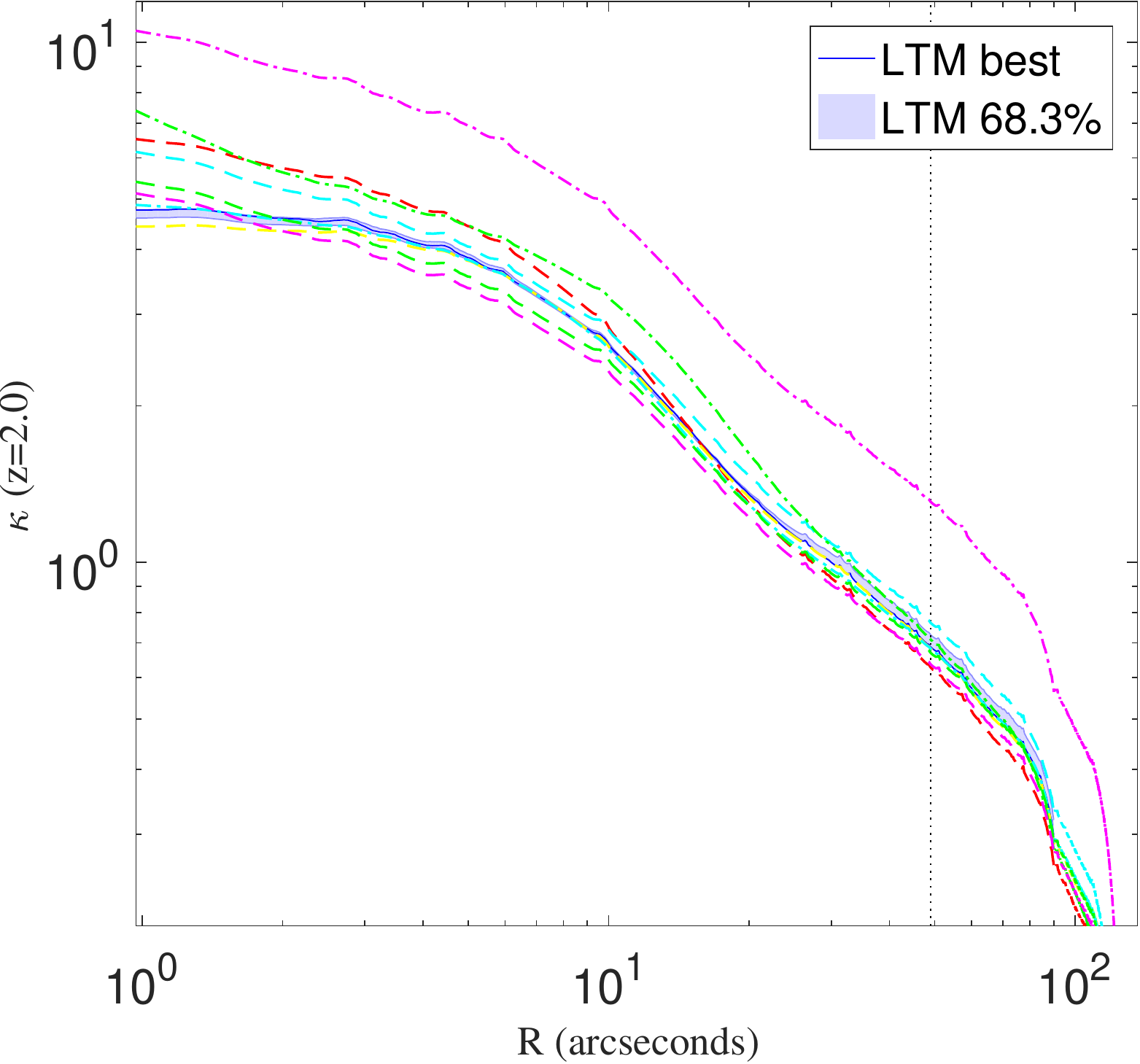} \hfill
 	\includegraphics[width=0.47\linewidth]{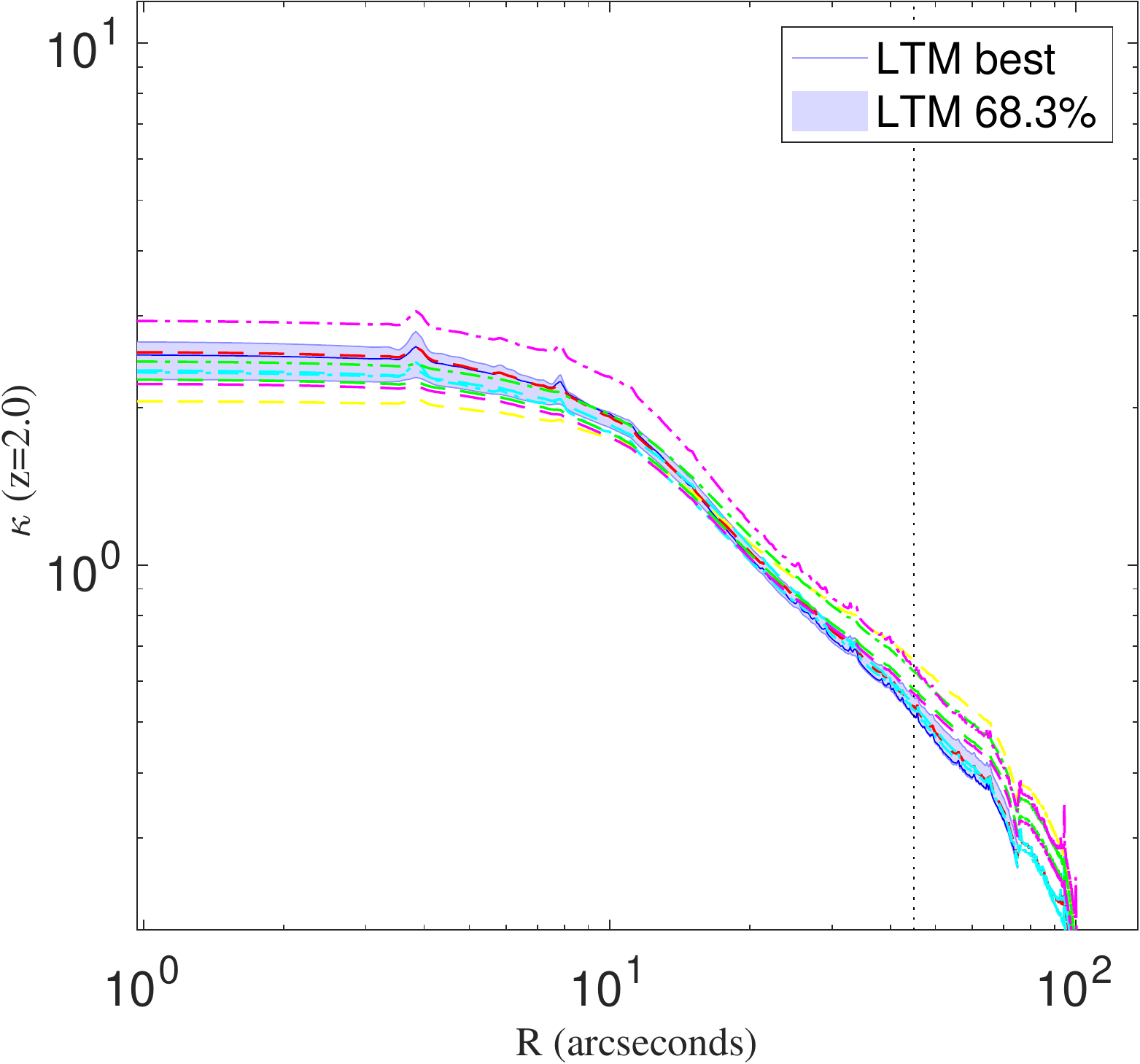}   
  
	\caption{Top panels - Map of the convergence, $\kappa$, representing the projected surface mass density in units of the critical density for lensing $\Sigma_{crit}$ and scaled a source redshift of $z_s\sim2.0$. Bottom panels - The corresponding, azimuthally averaged radial mass-density profile, and $1\sigma$ errors. The colored dashed lines show the range spanned by models considering different choices of photometric redshifts. The red and yellow lines are obtained when considering systems 2 and 3 as the main sources with fixed redshifts, respectively. Cyan, green and magenta dashed and dotted-dashed lines represent models considering a redshift value of $\pm10\%$,  $\pm25\%$ and $\pm50\%$ (respectively) of that used in the fiducial model (in blue). The black dashed vertical line sets the radius within which we have multiple images. Left column: MACS0308. Right column: PLCK G171.9}
	\label{kappa}
\end{figure*} 


\begin{figure*}
	\centering
	\includegraphics[width=0.495\linewidth]{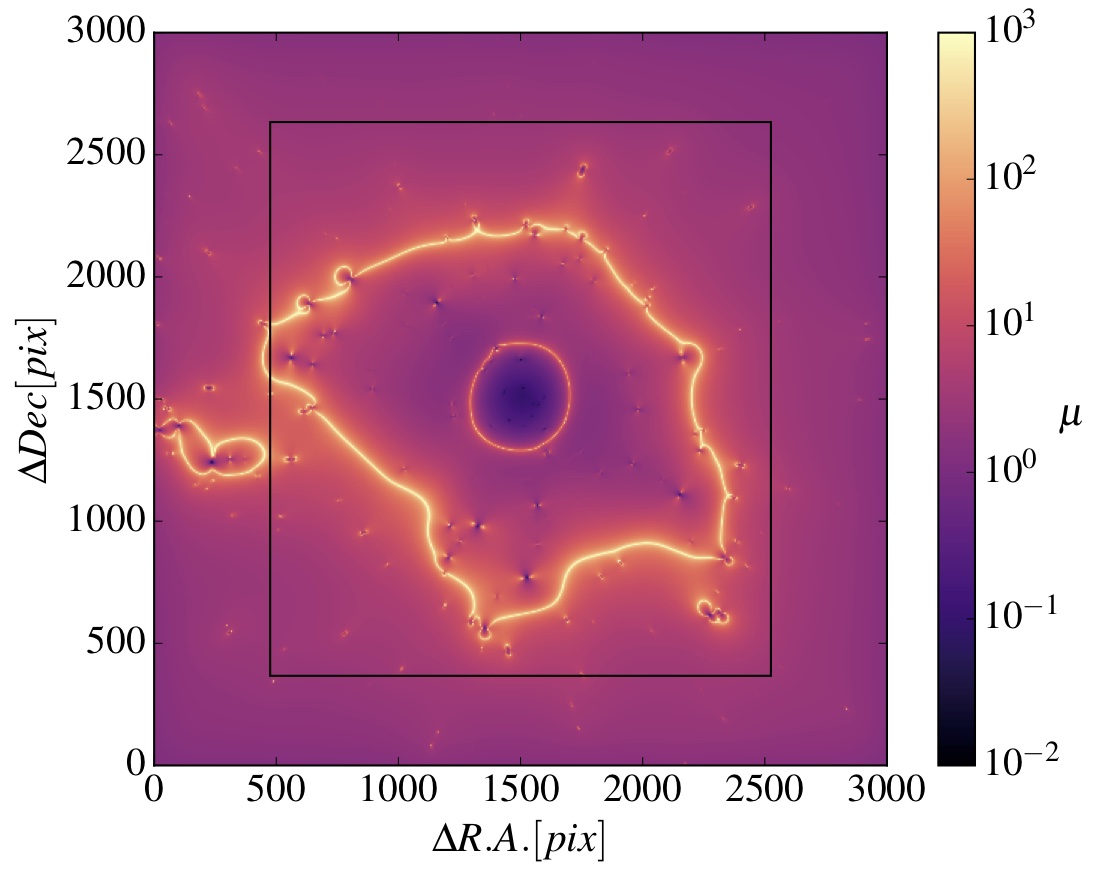} \hfill
	\includegraphics[width=0.495\linewidth]{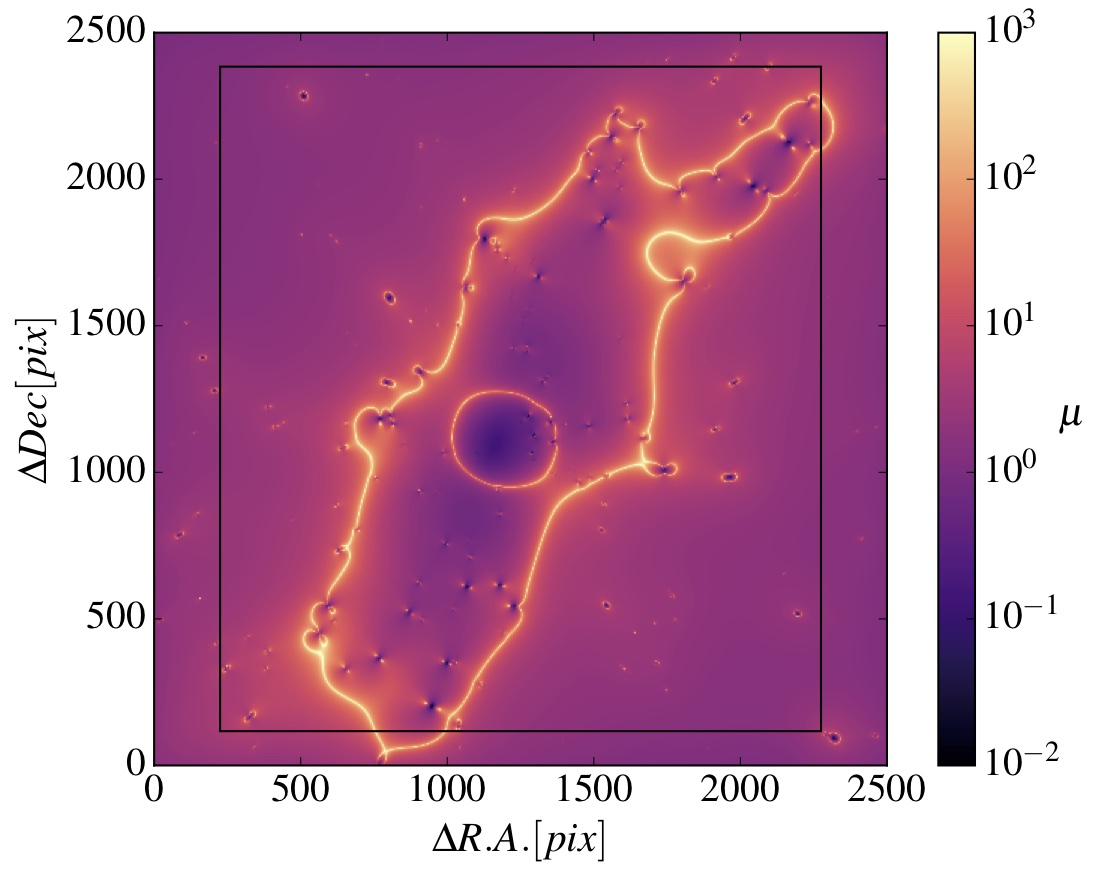}\\
	\includegraphics[width=0.4835\linewidth]{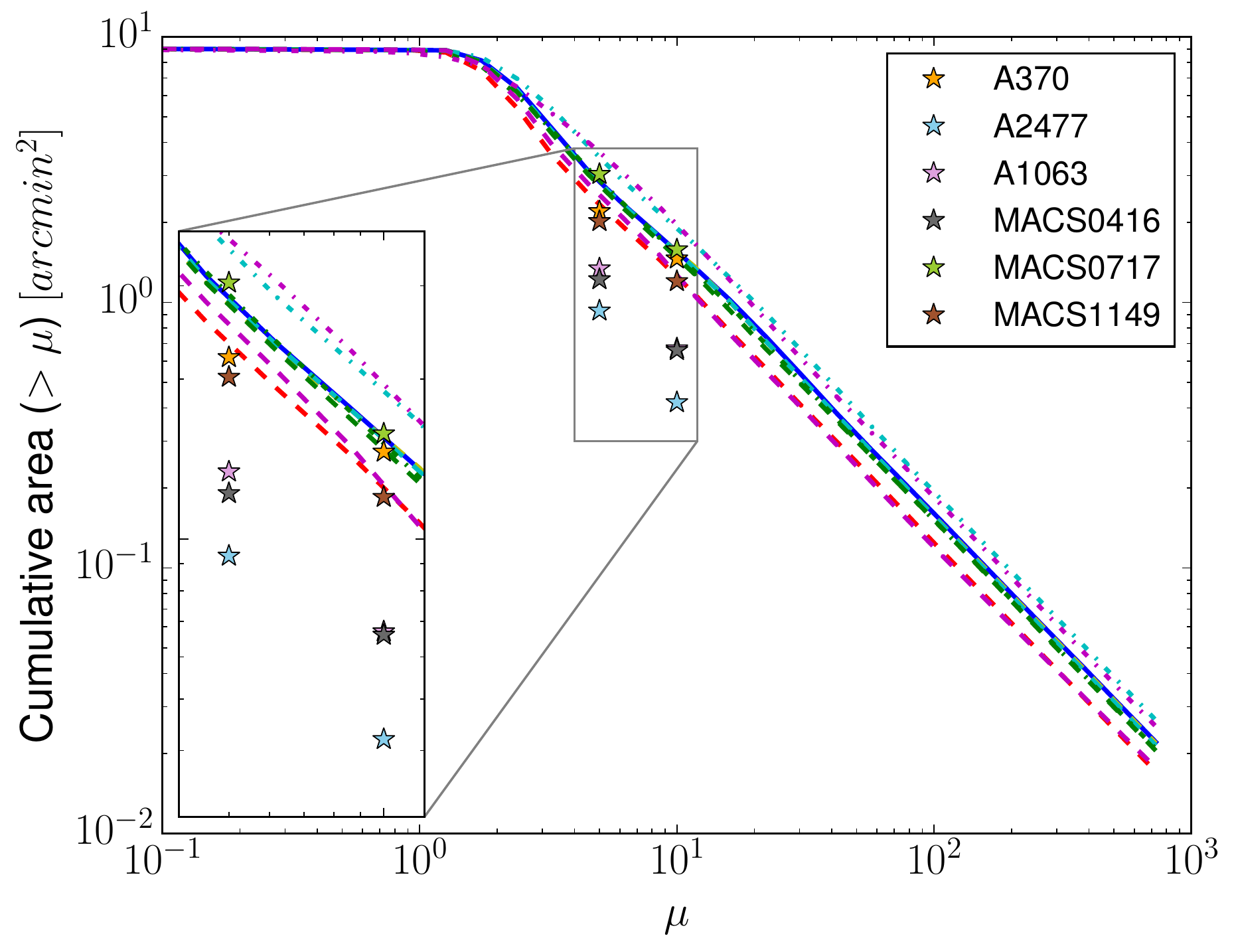} \hspace{0.2cm}
    \includegraphics[width=0.46\linewidth]{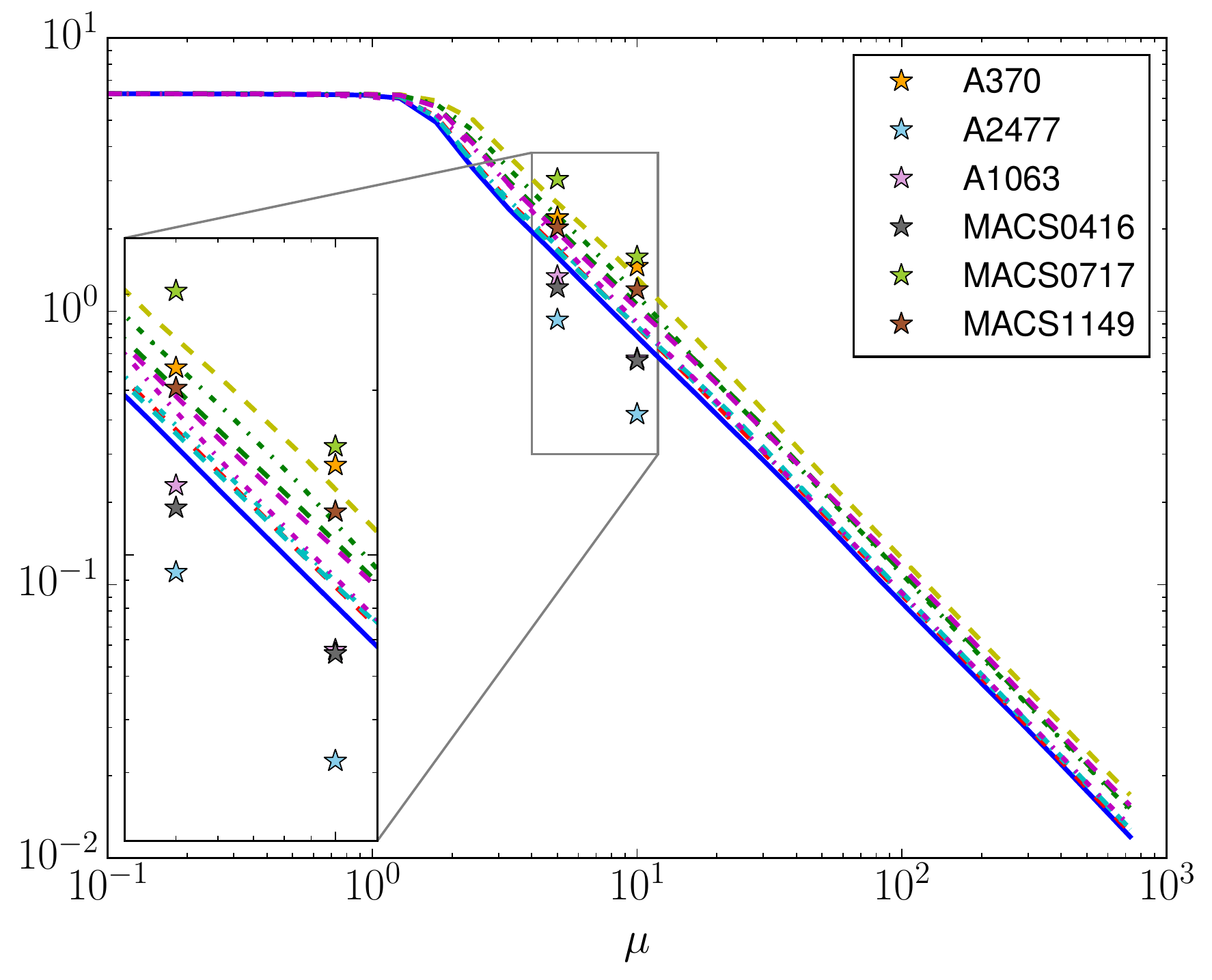} \hspace{0.5cm}
	\caption{Magnification maps for a source at $z_s=9.0$ from our best-fit models (upper row). The black rectangle indicates the WFC3/IR FOV. On the bottom row we show the corresponding cumulative area having a magnification higher than a given number for a source at $z_s=9.0$ (in blue). The colored dashed and dotted-dashed lines show the impact of our uncertainties due to the lack of spectroscopic redshifts with the same color code as in \autoref{kappa}. The cumulative areas ($\mu>5$ and $\mu>10$) for the \textit{Hubble Frontier Fields} clusters are also indicated as colored stars, computed from the submitted \textsc{zitrin-ltm-gauss} models. The $1\sigma$ errors are typically of the size of the star symbol. Left column: MACS0308. Right column: PLCK G171.9}
	\label{mag}
\end{figure*} 

\section{Results and discussion} \label{sec:results}
The surface mass-density distributions from our best-fit models for MACS0308 and PLCK G171.9, and their azimuthally averaged, radial 1D profiles, are shown in \autoref{kappa}. Our modeling of PLCK G171.9 reveals an elongated mass distribution in the NW-SE direction. This finding provides some further evidence for a merger scenario, suggested by previous X-ray studies \citep{Giacintucci2013}. In addition, PLCK G171.9 presents a particularly shallow inner mass profile, as a result of the two nearby BCGs. \\ For MACS0308, the mass distribution obtained from our lens modeling shows that it has a \textit{slight} elongation in the N-S axis, in agreement with that seen in radio observations \citep{parekh2017}, implying that the cluster is likely undergoing a merger. We note that the relative $1\sigma$ errors on the surface mass-density radial profiles are noticeably smaller than for PLCK 171.9, in part due to the smaller redshift range spanned by the multiple images behind PLCK 171.9.

Our SL analysis reveals that both MACS0308 and PLCK G171.9 are big lenses, with an effective Einstein radii (defined as $\theta_E=\sqrt[]{A/\pi}$ with $A$ the critical area, i.e., the area enclosed within the critical curves) of $\theta_E(z_s=2)= 33\pm 3\arcsec$ and $\theta_E(z_s=2)= 37\pm4\arcsec$, respectively. 
The mass enclosed within the critical curves is $ 2.5\pm0.4 \times 10^{14} M_{\odot}$ for MACS0308 and $ 1.3\pm0.2 \times 10^{14} M_{\odot}$ for PLCK G171.9. The uncertainties quoted here correspond to the typical errors found for these quantities, considering both statistical and systematic errors \citep{zitrin2015}.


Clusters of galaxies with large Einstein radii have proven to be ideal places to search for magnified high-redshift galaxies \citep[e.g.,][]{Zheng2012,Coe2013, Zitrin2014, Atek2015, Kawamata2016}.
In \autoref{mag} we present the magnification maps for a source at $z_s=9.0$ for both clusters, as well as a plot of the resulting cumulative area magnified above each magnification value, i.e., a plot of $A(>\mu)$ versus $\mu$, which can help to assess the strength of the lenses. We find that both clusters cover a fairly large area of high-magnification, $\sim 2.85(1.57)$ arcmin$^{2}$, for $\mu > 5$ to $\sim 1.56(0.81)$ arcmin$^{2}$ for $\mu=10$, for a source at a redshift of $z_s=9.0$ for MACS0308 (PLCK G171.9), which might provide in the future, interesting objects of the still elusive high-redshift population (i.e., those that are unreachable with current instruments). In \autoref{mag} we also mark the corresponding areas $A(\mu > 5)$ and $A(\mu > 10)$ for the \textit{Hubble Frontier Field} clusters \citep{Lotz2017}, computed from the \textsc{zitrin-ltm-gauss} models (in the respective, full area provided). Overall, the lensing strength of both RELICS clusters we analyze here is similar to that of the \textit{Hubble Frontier Field} clusters. Note that while the exact area in which the calculation is made affects the total normalization (the $A(\mu > 0)$ point), its effect diminishes as one goes to higher magnifications, as high magnifications are prominently induced only near the cluster center. Our current choice renders the strength estimation and its comparison to HFF clusters conservative, as our modeling field-of-view (FOV) -- about 6 and 10 arcmin$^2$ for PLCK G171.9 and MACS0308, respectively -- is equal to, or up to two times smaller than, that used for the HFF modeling (between 6 and 20 arcmin$^2$).


Recently, \citet{Salmon2017} presented the first sample of high-z candidates from RELICS, mainly at $z\sim 6 - 8$, by performing an independent analysis with two well-known photometric fitting-codes, BPZ and \textit{Easy and Accurate Z} \citep[EZ,][]{Brammer2008}. 
Lens models, provided following the RELICS observations, can sometimes yield further means to discriminate between photo-z solutions for high-z candidates, and, allow us to study the source intrinsic (i.e., demagnified) properties. Regarding the clusters considered in this work, \citet{Salmon2017} have identified six high-z ($z\sim6$) candidates within MACS0308's FOV and three for PLCK G171.9. 
Interestingly, MACS0308 hosts the third brightest high-z source from the RELICS sample \citep{Salmon2017}, MACS0308+26-0904 (J$\sim23.2$ AB), which is substantially sheared.

Our SL model predicts that MACS0308+26-0904 forms a multiply lensed system together with the second brightest high-z source within MACS0308's FOV, designated MACS0308+26-0438 (J$\sim24.6$ AB). This system is labelled c4 (see \autoref{table:0308}) and our SL model, together with the extensive photometric analysis by \citet{Salmon2017}, provide a redshift estimation of $z\sim6.4$. In addition, being two of the brightest sources in the RELICS high-z candidate sample, 
these two bright images are thus promising targets for follow-up spectroscopy. The third model-predicted counter-image is predicted near the cluster's center, and thus contaminated by the BCG's light (images c4.1 and c4.2 are radial images, and c4.3 is seen on the other side of the cluster as seen in \autoref{macs0308cc}). These images, as well as their reproduction by our SL model are shown in \autoref{stamps0308}.
We also examined another red image of a background source, labelled c4.3b in \autoref{table:0308}, close to our model's predicted location. This source has however an estimated photometric redshift $\mathrm{z_{phot}}\sim2.0$ (using BPZ), leading to the conclusion that c4.3 (MACS0308+26-0904) is likely the true counter image.

We present in \autoref{table:highzcan} the high-z candidates detected by \citet{Salmon2017}, following their notations, within MACS0308 and PLCK G171.9's FOV. For each high-z candidate we present a magnification estimate (and statistical uncertainty) from our best-fit model, as well as the absolute magnitude, $M_{uv}$, at $\lambda = 1500$ \AA. The absolute magnitude is computed from the UV continuum slope $f_{\lambda}\propto \lambda^\beta$ parametrization for galaxies \citep{Meurer1999}, obtained by a simple weighted least-squares fit using the four WFC3/IR bands (F105W, F125W, F140W, and F160W). The flux corresponding to the redshifted $\lambda = 1500$ \AA\ is then used to obtain the absolute magnitude, given by $M_{AB}=31.4-2.5\log_{10}(\mathrm{F_{nJy}})$.
As input we use the BPZ redshift estimate except in ambiguous cases where BPZ predicts a low redshift ($z\sim 1$), for which we then use the EZ estimate (as the scope is to characterize the intrinsic properties of high-z candidates). The resulting rest-frame UV luminosities (corrected for lensing magnifications) have a mean of $\mathrm{M_{uv}}\sim-19.45$ and standard deviation of 1.5.

Even though recent efforts have aimed at extensively targeting multiple images with ground-based follow-up spectroscopy within the \textit{Hubble Frontiers Fields} program for instance \citep{Mahler2018,Caminha2017}, most other galaxy clusters have, at best, only a few (if any) systems spectroscopically confirmed. Until more data becomes available, a main limitation of SL modeling lies therefore on the lack of spectroscopic redshifts for the background sources used as constraints (as well as the misidentification of multiple images). In order to accurately determine the mass distribution of the lens, both a reliable estimation of the cosmological distances between the observer and the source and the position of the multiple images in the image plane are crucial.
A lens model assuming only photometric redshifts can, for instance, under-predict the mass of cluster by up to $10\%$ within the Einstein radius \citep{Johnson2016}. More recently, \citet{Cerny2017} presented a SL analysis of the first five RELICS clusters and addressed the uncertainties introduced when considering photometrically estimated redshifts. The authors, using a parametric lens modeling algorithm, found that magnification is still constrained to better than $20\%$ in at least $80\%$ of the FOV, even when no spectroscopic redshifts were available. 

\begin{table*}
	\caption{High-z ($z \sim 6$) lensed candidates}            
	\label{table:highzcan}      
	\centering  
    {\renewcommand{\arraystretch}{1.6}
	\begin{tabular}{c c c c c c c c}        
		\hline\hline                 
		Galaxy ID\tablenotemark{a} & R.A. & Dec& $\mathrm{J_{125}}$ \tablenotemark{b}& $\mathrm{z_{phot}^{BPZ}}$\tablenotemark{c} & $\mathrm{z_{phot}^{EZ}}$\tablenotemark{d}& $\mu$ \tablenotemark{e}&  $M_{uv,1500}$\tablenotemark{f} \\  
		&[J2000]&[J2000]&[AB]&  &&&[AB]\\
		\hline          
		MACS0308+26-0904*   & 03:08:53.407 & +26:44:58.93 & $23.20 \pm 0.05$ &$6.3^{+0.1}_{-0.1}$& $6.4^{+0.2}_{-0.2}$& $16.99(19.11)^{+3.75}_{-2.12}$& $-20.43^{+0.48}_{-0.33}$\\ 
        MACS0308+26-0438*   & 03:08:57.189 & +26:45:48.37 & $24.64 \pm 0.09$ &$6.3^{+0.5}_{-0.3}$& $6.4^{+0.5}_{-0.3}$& $2.14(2.24)^{+0.07}_{-0.08}$&$-21.22^{+0.32}_{-0.31}$ \\
        MACS0308+26-0249   & 03:08:57.025 & +26:46:07.15 & $23.75 \pm 0.11$ &$5.6^{+0.2}_{-0.5}$& $1.0^{+0.3}_{-0.1}$& $3.07(3.00)^{+0.09}_{-0.09}$&$-20.66^{+0.31}_{-0.31}$\\
        MACS0308+26-0991   & 03:08:57.197 & +26:44:41.96 & $25.93 \pm 0.22$ &$5.4^{+0.4}_{-4.8}$& $6.0^{+0.3}_{-1.3}$& $11.93(10.66)^{+1.24}_{-0.94}$&$-17.80^{+0.37}_{-0.87}$\\ 
        MACS0308+26-0184   & 03:08:57.288 & +26:46:19.27 & $26.66 \pm 0.29$ &$5.3^{+0.4}_{-1.0}$& $5.7^{+0.4}_{-0.9}$& $3.14(3.17)^{+0.11}_{-0.08}$&$-18.87^{+0.31}_{-0.35}$\\ 
        MACS0308+26-0575  & 03:08:51.383 & +26:45:36.89 & $26.93 \pm 0.36$ &$0.9^{+5.1}_{-0.4}$& $5.8^{+0.5}_{-5.1}$& $10.58(13.27)^{+3.55}_{-2.69}$&$-17.12^{+0.71}_{-0.91}$\\ 
        \hline  
        PLCKG171-40-0130  & 03:12:54.204 & +08:23:03.88 & $23.93\pm 0.12$ &$5.7^{+0.3}_{-5.1}$ &$5.8^{+0.4}_{-1.1}$& $4.54(6.05)^{+1.30}_{-0.99}$& $-20.77^{+0.63}_{-0.91}$\\
        PLCKG171-40-0355  & 03:12:57.268 & +08:22:35.47 & $25.41\pm 0.19$ &$1.0^{+5.0}_{-0.3}$ &$5.9^{+0.4}_{-0.8}$& $1.96(2.32)^{+0.14}_{-0.16}$&$-20.53^{+0.35}_{-0.34}$\\
        PLCKG171-40-0738  & 03:12:57.520 & +08:21:51.10 & $25.35\pm 0.15$ &$5.5^{+0.3}_{-4.8}$ &$5.7^{+0.4}_{-4.7}$& $21.32(32.42)^{+9.93}_{-12.01}$ &$-17.62^{+1.16}_{-1.05}$\\
		\hline\hline                                
	\end{tabular}}
      \tablecomments{}
      \tablenotetext{1}{Galaxy ID, following \citet{Salmon2017} notations.The first two high-z candidates, indicated with an asterisk, correspond to the multiply imaged candidates c4.3 and c4.1 in \autoref{table:0308}, respectively}
     \tablenotetext{2}{Apparent magnitude in the F125W band.}
  \tablenotetext{3}{Redshift estimation based on the BPZ pipeline along with their $1\sigma$ uncertainties.}
 \tablenotetext{4}{Redshift estimation based on the EZ pipeline along with their $1\sigma$ uncertainties.}
  \tablenotetext{5}{Magnification estimates (at the respective source redshift) from our best-fit model, the average (computed from 100 random MCMC models, in parenthesis) and statistical uncertainty. The best-fit value is the one used for all relevant computations.}
  \tablenotetext{6}{Absolute magnitude, $M_{uv}$, at $\lambda = 1500$ \AA\ for which the errors have been propagated from the photometric and magnification uncertainties.}
\end{table*}

As neither cluster analyzed here has any multiple image spectroscopically confirmed yet, we have further assessed the robustness of our results (regarding the mass distribution and magnification) by running several models with different combinations of fixed source redshifts, or by varying the main source redshift (which was fixed in the fiducial model) by $\pm10\%$, $\pm25\%$ and $\pm50\%$. The resulting, azimuthally averaged mass-density profiles and the strength of the lenses, presented as colored dashed lines, are compared to the best-fit model in the bottom panels of \autoref{kappa} and in \autoref{mag}, respectively. We find that, as expected, different combinations of photometric redshifts yield different estimations for the mass profiles, particularly in the inner regions. The mass profiles typically agree within $2\sigma-3\sigma$, where for the most extreme cases in which we vary the main source redshift, they can differ by $\sim50\%$. The Einstein radii inferred from each trial model agree to within $10\%$, and only provide a larger Einstein radius when other redshift combinations are assumed (so that the Einstein radius we quote is on the conservative side). Finally, we have also assessed the impact of redshift uncertainties on the determination of the strength of the lens, in terms of the cumulative magnification seen in \autoref{mag}. While these uncertainties do not have a significant impact for the total magnified area, i.e., the area with magnifications above $\mu'$ where $\mu'$ is small ($\mu' \lesssim1-2$), higher-magnification regions can suffer up to a $60\%$ bias, and, in the case of PLCK G171.9, only increasing the strength of the lens when considering other redshift combinations (so also our magnification estimate appears to be conservative). Overall we find that magnification at $z=9$ is constrained to better than $20\%$ in at least $70\%$ and $50\%$ of the modeled FOV when overestimating or underestimating the fixed main source redshift by $10-25\%$, respectively; similar to the findings \citet{Cerny2017} obtained with a fully parametric modeling technique. By changing the value of the fixed main source redshift by $50\%$, the magnification is constrained to better than $30\%$ in around $60\%$ of the modeled FOV.
These typical errors include the uncertainties from redshift or cosmological distances and thus, in part, the mass-sheet degeneracy \citep{Falco1985,Liesenborgs2012}. \\
Other sources of systematic errors, beyond the scope of this study, should also be taken into account as they can introduce a supplementary bias on magnification such as the presence of non-correlated line-of-sight haloes \citep{Daloisio2014, Chirivi2017, Williams2017}, the assumption that light traces mass \citep{Harvey2016}, different modeling choices \citep{Limousin2016}, the triaxial shape of clusters \citep{Giocoli2014, Sereno2015} or uncertainties regarding the cosmological parameters \citep{Bayliss2015}. For a more general discussion of systematics in SL analysis see \citet{zitrin2015, Meneghetti2017} and \citet{Priewe2017}. 

Our SL models are made ublicly available through the MAST archive as RELICS high-level science products. We supply deflection fields, surface mass density ($\kappa$) maps, and magnification maps for different redshifts, including also a subset of a hundred random models from the MC so that errors can be calculated. An online magnification calculator will also be available\textsuperscript{\ref{mast}} for fast magnification estimates.  

We note that one should be generally cautious when using SL modeling outputs (such as convergence, magnification, etc.) beyond the SL regime where multiple images are seen (roughly twice the Einstein radius), so that the lens models should be considered extrapolations beyond this limit. In addition, in our case, the smoothing and other interpolations used in our methodology often introduce some boundary artifacts, mainly near the edges of the modeled FOV, of which the user should be aware.

\section{Summary}\label{sec:summary}
In this work we have analyzed two massive, non-relaxed clusters from the RELICS cluster sample, MACS J0308.9+2645 and PLCK G171.9-40.7, that are X-ray and SZ-selected, respectively.  
We presented the first SL analysis of both clusters, adopting a Light-Traces-Mass methodology and uncovering several multiply imaged galaxies in each. Our analysis has revealed that both MACS0308 and PLCK G171.9 are prominent lenses with large Einstein radii, hosting some promising $z\sim 6$ candidates \citep{Salmon2017}. Both clusters are therefore viable targets for spectroscopic follow-up, both to obtain spectroscopic redshifts of the multiple-image systems presented in this work to refine the lens model, as well as for studying high-z candidates. Particularly, MACS0308 hosts a particularly bright (J$\sim23.2-24.6$ AB) multiply imaged high-z candidate useful for spectroscopic follow-up, with both photometric analysis \citep{Salmon2017} and our lens model agreeing on its high-redshift nature at $z\sim6.4$. This source is amongst the brightest high-z candidates found in all of RELICS.

Our lens models, as well as magnification maps, are made publicly available through the MAST archive\textsuperscript{\ref{mast}}.
Massive galaxy clusters with large Einstein radii are also statistically interesting, enabling to determine the high-mass end of the halo distribution and probe theoretical $\Lambda \mathrm{CDM}$ predictions for structure formation and evolution \citep{Oguri2009,Redlich2014}. In that sense, it should also be noted that the selection function plays a crucial role and should be taken into account in such comparisons. Moreover, such studies suffer from a modest sample size, as only several to a dozen galaxy clusters with $\theta_E > 30"$ are known to date \citep[see for instance][]{broadhurst2005,Limousin2007,Zitrin2009b, Zitrin2017, Richard2010, Postman2012, Cerny2017}.
In this work we present and add to the list two new massive clusters with such large Einstein radii. These will be potentially complemented with other large lensing clusters from the RELICS sample that are now being analyzed.

To estimate leading uncertainties in our models, we explored possible biases due to the lack of spectroscopic redshifts for the multiply imaged systems. We find that, while a refinement of the models is warranted when spectroscopic data becomes available, our SL models yield a robust measurement (within $10\%$) of the Einstein radius for both clusters, taking into account both the statistical, and systematic errors arising from the lack of spectroscopic data. This translates into a similar $\sim15\%$ for the enclosed, Einstein mass. For the mass distribution, and in particular the 1D radial mass-density profile, we find that different redshift combinations agree to within 2 or 3-$\sigma$. Magnification estimates appear to be more susceptible to the redshift combinations, and can reach a $60\%$ discrepancy in areas with high-magnification. These values should thus be used more cautiously, taking into account the relevant systematic uncertainties. Additional explicit biases in the quantities derived in this work can be further assessed by comparing our results to those from other independent lens modeling techniques \citep{Meneghetti2017,Bouwens2017}.

Together with the spectacular imaging data provided by the RELICS project, SL models of the whole cluster sample \citep[see also][]{Cerny2017, Cibirka2018} will help divulge exciting high-redshift candidates \citep{Salmon2018} in time for \textit{James Webb Space Telescope} spectroscopic follow-up. 

\acknowledgments 
We thank the anonymous referee for her/his useful comments that helped to improve this work. This work is based on observations taken by the RELICS Treasury Program (GO-14096) with the NASA/ESA HST. Program GO-14096 is supported by NASA through a grant from the Space Telescope Science Institute, which is operated by the Association of Universities for Research in Astronomy, Inc., under NASA contract NAS5-26555. This work was performed in part under the auspices of the U.S. Department of Energy by Lawrence Livermore National Laboratory under Contract DE-AC52-07NA27344.
RCL acknowledges support from an Australian Research Council Discovery Early Career Researcher Award (DE180101240).

%

\vspace{5mm}



\bibliographystyle{apj}
\bibliography{test}
\appendix

\begin{appendices}
 \section{Reproduction of the multiple images}
 
 \begin{figure*}[h!]
	\centering
	\includegraphics[width=0.9\linewidth]{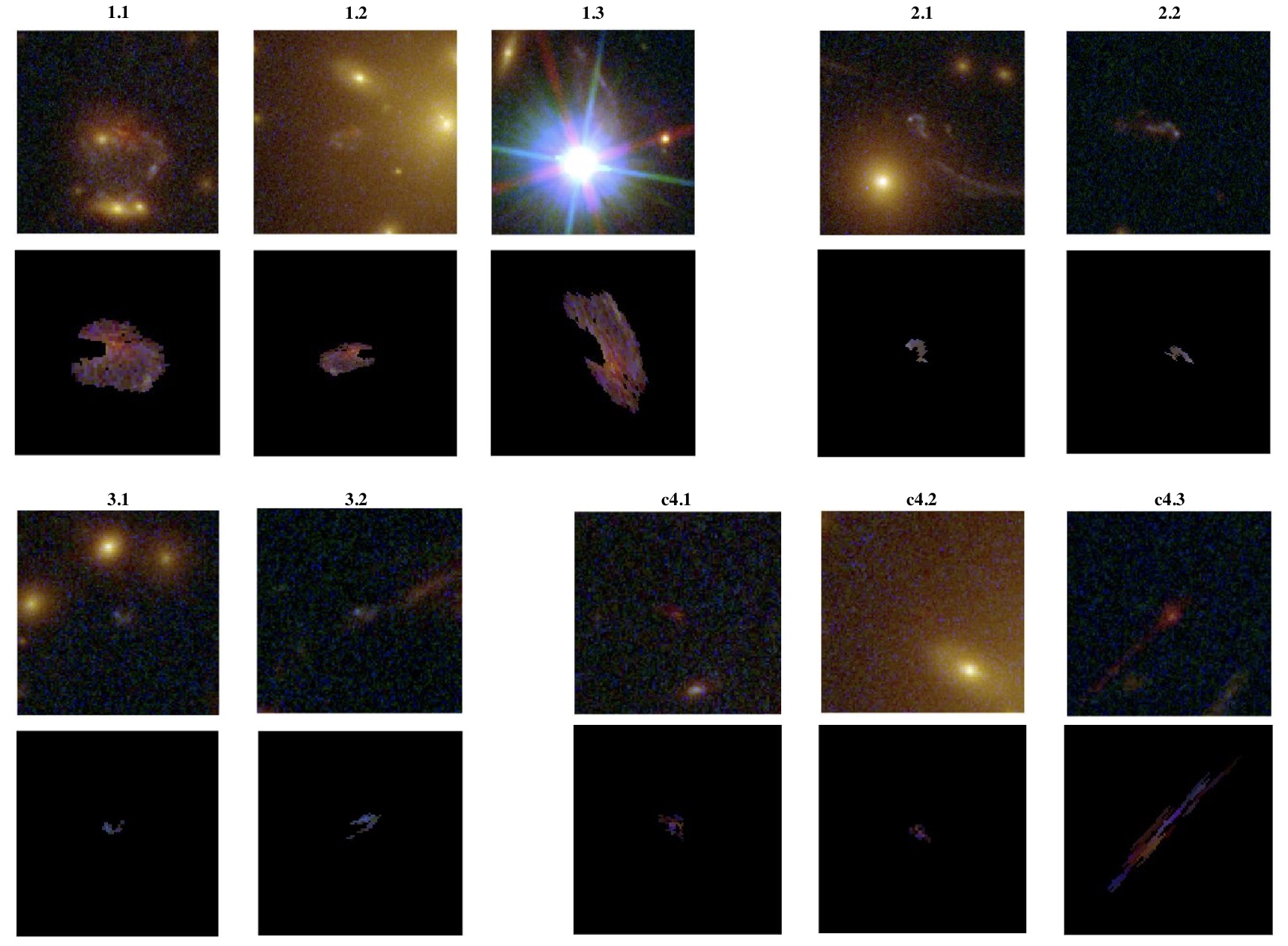} \\
	\caption{Reproduction of multiple images by our best-fit model for MACS0308. For each image, we de-lens the first image of the system to the source plane and back to the image plane to compare to the other images of that system. The orientation and internal details of the model-predicted images (bottom rows) are similar to those of the observed images (upper rows). }
	\label{stamps0308}
\end{figure*} 

 \begin{figure*}[h!]
	\centering
	\includegraphics[width=0.95\linewidth]{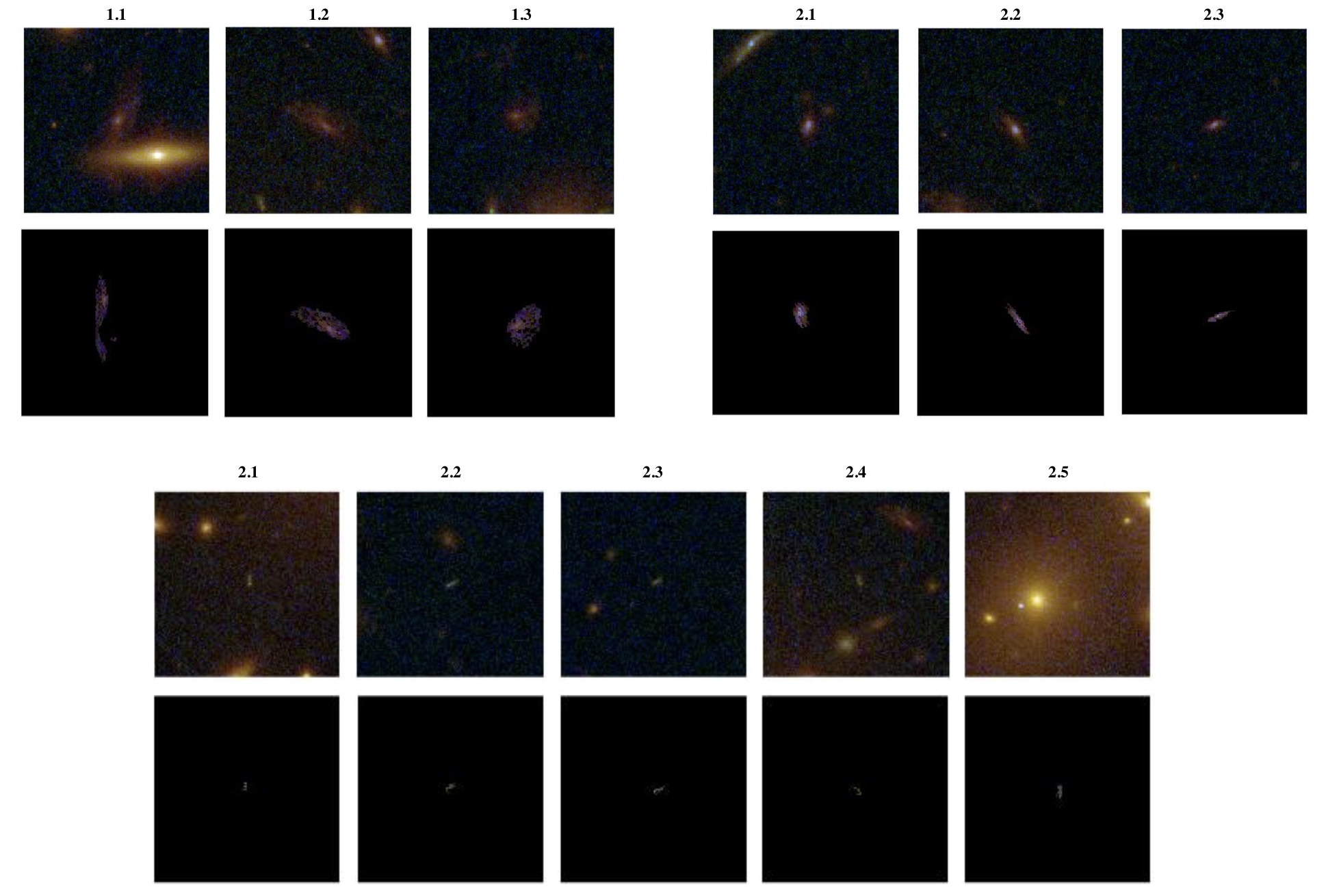}
	\caption{Reproduction of multiple images by our best-fit model for PLCK G171.9. Produced similarly to \autoref{stamps0308}.}
	\label{stamps171}
\end{figure*} 

\end{appendices}


\end{document}